# Atomic-Layer-Controlled Magnetic Orders in MnBi$_2$Te$_4$-Bi$_2$Te$_3$ Topological Heterostructures

Xiong Yao[1, 2, #, *], Qirui Cui[1, 3, #, †], Zengle Huang[4], Xiaoyu Yuan[4], Hee Taek Yi[4], Deepti Jain[4],

Kim Kisslinger[5], Myung-Geun Han[6], Weida Wu[4], Hongxin Yang[3, *], and Seongshik Oh[2, *]

[1]Ningbo Institute of Materials Technology and Engineering, Chinese Academy of Sciences,

Ningbo 315201, China

[2]Center for Quantum Materials Synthesis and Department of Physics & Astronomy, Rutgers, The

State University of New Jersey, Piscataway, New Jersey 08854, United States

[3]Center for Quantum Matter, School of Physics, Zhejiang University, Hangzhou 310058, China

[4]Department of Physics & Astronomy, Rutgers, The State University of New Jersey, Piscataway,

New Jersey 08854, United States

[5]Center for Functional Nanomaterials, Brookhaven National Laboratory, Upton, New York

11973, United States

[6]Condensed Matter Physics and Materials Science, Brookhaven National Laboratory, Upton,

New York 11973, United States

[#]Xiong Yao and Qirui Cui contributed equally to this work.

†Present address: Department of Applied Physics, School of Engineering Sciences, KTH Royal

Institute of Technology, AlbaNova University Center, 10691 Stockholm, Sweden

Email: yaoxiong@nimte.ac.cn, hongxin.yang@zju.edu.cn, ohsean@physics.rutgers.edu

ABSTRACT

The natural van der Waals superlattice $MnBi_2Te_4$-$(Bi_2Te_3)_m$ provides an optimal platform to combine topology and magnetism in one system with minimal structural disorder. Here, we show that this system can harbor both ferromagnetic (FM) and antiferromagnetic (AFM) orders and that these magnetic orders can be controlled in two different ways by either varying the Mn-Mn distance while keeping the $Bi_2Te_3$/$MnBi_2Te_4$ ratio constant or vice versa. We achieve this by creating atomically engineered sandwich structures composed of $Bi_2Te_3$ and $MnBi_2Te_4$ layers. We show that the AFM order is exclusively determined by the Mn-Mn distance whereas the FM order depends only on the overall $Bi_2Te_3$/$MnBi_2Te_4$ ratio regardless of the distance between the $MnBi_2Te_4$ layers. Our results shed light on the origins of the AFM and FM orders and provide insights into how to manipulate magnetic orders not only for the $MnBi_2Te_4$-$Bi_2Te_3$ system but also for other magneto-topological materials.

Intrinsic magnetic topological insulators (MTIs) such as the $MnBi_2Te_4$-$(Bi_2Te_3)_m$ (m = 0, 1, 2, ...) compounds, which are composed of a natural stacking of stoichiometric van der Waals units, incorporate topological states and magnetism in a manner that introduces significantly less structural disorder than magnetic doping[1-18]. The experimental demonstration of quantum anomalous Hall effect (QAHE) and robust axion insulator state in the $MnBi_2Te_4$ compound[2, 3, 19] highlights its potentials for topological spintronics. The $MnBi_2Te_4$ layers are ferromagnetically coupled within each septuple layer (SL) and antiferromagnetically coupled between SLs in A-type configuration[20].

In the $MnBi_2Te_4$-$(Bi_2Te_3)_m$ bulk crystals, the antiferromagnetic (AFM) interlayer coupling becomes weaker and eventually turns into ferromagnetic (FM) as non-magnetic $Bi_2Te_3$ quintuple layers (QLs) are intercalated between the $MnBi_2Te_4$ layers[21-24]. Along this line, people successfully synthesized $MnBi_4Te_7$ (i.e. m = 1), $MnBi_6Te_{10}$ (m = 2) and $MnBi_8Te_{13}$ (m = 3) bulk crystals and found coexistence of AFM and FM orders for m = 1 and 2 but FM phase for m = 3[21-23, 25, 26], as summarized in Figure 1a. However, introduction of the $Bi_2Te_3$ extra spacer layers in the $MnBi_2Te_4$-$(Bi_2Te_3)_m$ system increases not only the interlayer distance between the two neighboring $MnBi_2Te_4$ layers (Mn-Mn distance) but also the $Bi_2Te_3/MnBi_2Te_4$ ratio. The latter effect, which dilutes the overall magnetic ion concentration, possibly plays a non-negligible role in the evolution of magnetic orders, just like the Cr concentration-dependent magnetism in Cr-doped $(Bi,Sb)_2Te_3$[27-29]. Accordingly, the exact role of $Bi_2Te_3/MnBi_2Te_4$ ratio apart from the influence of Mn-Mn distance in these $MnBi_2Te_4$-$(Bi_2Te_3)_m$ compounds is unclear [21-24]. Moreover, with the superlattice structure in these bulk crystals, it would not be possible to independently control Mn-Mn distance and $Bi_2Te_3/MnBi_2Te_4$ ratio. On the other hand, recent studies report that magnetic defects such as Mn antisites can induce ferromagnetism in bulk crystals of $MnBi_4Te_7$, $MnBi_6Te_{10}$, and $MnSb_2Te_4$[30-38], the sister compound of $MnBi_2Te_4$. In order to precisely engineer the magnetic properties in such complex magneto-topological materials, clear understanding of all these effects, especially the exact roles of $Bi_2Te_3/MnBi_2Te_4$ ratio and the Mn-Mn distance, is essential. The layer-by-layer molecular beam epitaxy (MBE) technique allows us to independently control those two effects at atomic level through an artificial layering scheme[39], providing a strategy to disentangle Mn-Mn distance and $Bi_2Te_3/MnBi_2Te_4$ ratio regarding their effects on the magnetic orders (both AFM and FM) in this system, which has never been explored previously.

**Independent Control of $Bi_2Te_3/MnBi_2Te_4$ ratio and Mn-Mn distance.** In order to investigate the effect of Mn-Mn distance and $Bi_2Te_3/MnBi_2Te_4$ ratio on the magnetic properties of the $MnBi_2Te_4$-$Bi_2Te_3$ heterostructures separately, we designed and grew two sets of $MnBi_2Te_4$-$Bi_2Te_3$ heterostructures (Figure 1b and c), using our atomic-layer-by-layer molecular beam epitaxy (MBE) technique [10, 38, 40-44]. In configuration (b), we added n QL $Bi_2Te_3$ layers on each side of a single slab of $MnBi_2Te_4$-$Bi_2Te_3$-$MnBi_2Te_4$ structure while keep the Mn-Mn distance constant. On the other hand, in configuration (c), the $Bi_2Te_3/MnBi_2Te_4$ ratio (total $Bi_2Te_3$ thickness) is fixed while the Mn-Mn distance is varied. It is important to note that all these samples are not a superlattice (which is a repetition of a unit cell structure), and rather they are a single slab heterostructure. As shown in Figure 1d, we achieved independent control of $Bi_2Te_3/MnBi_2Te_4$ ratio and the Mn-Mn distance in configuration (b) and (c), respectively, while both of these two effects vary together in the $MnBi_2Te_4$-$(Bi_2Te_3)_m$ bulk crystals (which are in superlattice structure). Figure 1e shows the typical high-angle annular dark-field scanning transmission electron microscopy (HAADF-STEM) image of a $MnBi_2Te_4$-$Bi_2Te_3$ heterostructure, which exhibits well-defined QL and SL structures of $Bi_2Te_3$ and $MnBi_2Te_4$ layers, respectively, with sharp van der Waals gaps between two adjacent layers. The sample quality was further confirmed by reflection high-energy electron diffraction (RHEED) patterns, as shown in Figure S1. It is notable that despite the well-defined QL and SL structures, stacking of $MnBi_2Te_4$ and $Bi_2Te_3$ layers is inherently non-uniform at the macroscopic scale for all MBE-grown films. Nonetheless, since transport measurements reflect the average properties of the entire sample and all our samples were grown under identical conditions, the non-uniform stacking is neither a variable factor in our study nor does it impact the conclusions in the subsequent sections.

**The origin of coexisting AFM and FM orders.** We found that AFM and FM orders coexist in many of our $MnBi_2Te_4$-$Bi_2Te_3$ heterostructures as shown in Figures 3 and 4. So, first, we investigate their origin in Figure 2, by combining scanning tunneling microscopy (STM) and first principles calculations. Previously it was reported that magnetic defects like Mn antisites or Mn-Bi site mixing can enhance FM order in $MnSb_2Te_4$, $MnBi_4Te_7$ and $MnBi_6Te_{10}$[30-34]. To determine the distribution of magnetic defects in our samples, we mapped out the $Mn_{Bi}$ antisite defects in the n = 0 sample (configuration (b)) by STM, as shown in Figure 2a. The $Mn_{Bi}$ antisites appear as dark defects in the marked triangles. The density of $Mn_{Bi}$ antisites are determined as 2.1% on average by counting the defects in the measured region in Figure 2a. On the other hand, for n = 4 (Figure 2b), the Mn density (~0.2%) is ten times smaller than the value in Figure 2a, strongly suggesting that the $Mn_{Bi}$ antisite defects are mostly confined within the $MnBi_2Te_4$ layer and the interlayer Mn diffusion is minuscule compared with the Mn density in the $MnBi_2Te_4$ layer.

To give a further insight into the relationship between $Mn_{Bi}$ magnetic defects and the FM order, we conducted first-principles calculations of the $MnBi_2Te_4$-$Bi_2Te_3$ heterostructure where some Bi atoms are substituted by Mn atoms for simulating $Mn_{Bi}$ defects, as shown in the left panel of Figure 2c. The ground state of the $MnBi_2Te_4$-$Bi_2Te_3$ heterostructure without Mn antisite defects was determined as AFM (right panel of Figure 2c). However, as more Bi sites are replaced by Mn atoms, the energy difference between AFM and FM coupling quickly narrows to nearly 0 at the critical Mn doping level of 3.57%, and FM coupling becomes the ground state, then reaching its peak strength at 7.14% of Mn before decreasing with further Mn. It is worth noting that the AFM coupling would become the ground state again above 14.28% of Mn doping according to the trend in Figure 2c, which is consistent with the observations in MnTe intercalated $MnBi_2Te_4$ superlattices[41]. Even though Figure 2c only gives one possible configuration for Mn substitution, this suggests that the magnetic ground state can be easily switched between AFM and FM state depending on the level of Mn defect densities. As observed in Figure 2a, the distribution of Mn defect concentration is inhomogeneous in our films, suggesting that the Mn defect density could be lower than the critical density of AFM-to-FM transition at Mn poor regions but higher than that value at Mn clustered regions. In other words, inhomogeneous Mn defect densities can naturally explain the coexistence of FM and AFM orders in our $MnBi_2Te_4$-$Bi_2Te_3$ heterostructures.

**The effect of $Bi_2Te_3$/$MnBi_2Te_4$ ratio on AFM and FM orders.** In configuration (b), the top and bottom n QL $Bi_2Te_3$ layers increase only the $Bi_2Te_3$/$MnBi_2Te_4$ ratio, while the Mn-Mn distance is fixed. To investigate the exact role of these outer extension layers, we performed systematic longitudinal and Hall resistance measurements on samples with varying n values. Figure 3a exhibits the corresponding magnetic-field-dependent longitudinal resistance for all the samples measured at 2 K. There are several peculiar features in Figure 3a. First, in samples n = 0, 1, and 2, we can clearly observe two symmetric shoulder peaks at around 3 T. This shoulder feature is commonly observed in $MnBi_2Te_4$ and is related to magnetic-field-driven spin-flopping process in antiferromagnet[3], suggesting the presence of AFM order. As marked by the dash lines in Figure 3a, the spin-flop magnetic field remains unchanged for samples n = 0, 1 and 2, implying the energy scale or strength of AFM order remains unaffected by increasing n values. The intensity of the shoulder features gradually reduces as n increases and then completely vanishes in samples n = 3, 4 and 5, likely due to the smearing effect from increased conduction contribution of additional $Bi_2Te_3$ extension layers. Second, all the curves exhibit clear negative magnetoresistance (MR) peaks around zero field with hysteresis in samples n = 0 to n = 4 (shown in Figure S2), which are generally observed in various FM materials and regarded as features of FM order [2, 40, 45]. Figure S2 gives the enlarged plot of Figure 3a at low magnetic fields. With increasing $Bi_2Te_3$ extension layers, the characteristic magnetic field of FM hysteresis peaks gradually reduces and eventually approaches zero, indicating weakening FM order.

Hall resistance is another probe to detect the magnetic signatures in thin film samples. Figure 3b presents the temperature-dependent Hall resistance measured under zero magnetic field while cooling down for all the samples. All the $R_{xy}$ vs T curves show clear transitions at low temperatures, which is direct evidence of spontaneous magnetization induced by FM order. The transition temperatures show a monotonic decrease with increasing n values, implying reduced FM order. We conducted Hall resistance loop measurements on all the samples at 2 K, as shown in Figure 3c. Notably, they all exhibit clear hysteresis loops, indicating net magnetization resulting from ferromagnetism. As the number of $Bi_2Te_3$ extension layers n increases, the size of hysteresis loops monotonically shrinks in Figure 3c, reflecting weakening FM order. The coercive fields $H_C$ and the magnetic transition temperature $T_C$ follow very similar trend with increasing n values as shown in Figure 3d, suggesting that both quantities originate from the same (FM) origin. Summarizing the results of Figure 3, we can conclude that AFM and FM orders coexist in these samples and that the added $Bi_2Te_3$ extension layers, which increase the $Bi_2Te_3/MnBi_2Te_4$ ratio, significantly weakens FM order while keeping AFM order almost unaffected.

**The effect of Mn-Mn distance on AFM and FM orders.** In order to see how the Mn-Mn distance alone affects the AFM and FM orders, we implemented heterostructures in configuration (c) (as shown in Figure 1c and Figure 4a): this keeps the $Bi_2Te_3$/$MnBi_2Te_4$ ratio (also the overall composition) the same as that of the "n = 3" sample in Figure 1b. The corresponding longitudinal and Hall resistance are shown in Figure 4b-e. Interestingly, from Figure 4e we find that the spin-flop feature at around 3 T in the $R_{xx}$ vs $\mu_0H$ data, characteristic of the AFM order in Figure 3a, emerges when the Mn-Mn distance is reduced to "d = 0" from "d = 1" but disappear as the Mn-Mn distance increases. It is worth to mention that the vanishing spin-flop feature here are not likely due to the smearing effect from $Bi_2Te_3$ layers, as the total thickness of $Bi_2Te_3$ (also the conduction contribution from $Bi_2Te_3$) are fixed for all the samples in Figure 4a. This observation indicates that the AFM interlayer coupling in d = 0 is stronger than other samples, owing to the absence of $Bi_2Te_3$ spacer layers in the middle. Previous studies on $MnBi_2Te_4$-$(Bi_2Te_3)_m$ bulk crystals also suggest that the AFM order can be controlled by $Bi_2Te_3$ spacer layers, as illustrated by the decreasing $T_N$ in Figure 1a[21, 22, 24]. However, in the superlattice-like bulk crystal case, the effect of Mn-Mn distance is mixed with the impact of variable $Bi_2Te_3$/$MnBi_2Te_4$ ratio. Now we clarify that the Mn-Mn distance plays a dominating role in manipulating the AFM order in $MnBi_2Te_4$-$Bi_2Te_3$ systems, disentangling these two effects through the atomic-layer-controlled MBE growth, in a single-slab configuration.

On the other hand, Figures 4b-d show that the strength of FM order, as measured by $H_C$ and $T_C$, remains little affected by the Mn-Mn distance. With increasing Mn-Mn distance, the variation of both $H_C$ and $T_C$ are limited to fluctuation level and not exhibiting any observable trend. Such behavior is also in stark contrast with the observations in $(Cr,Bi,Sb)_2Te_3$ system, where the FM order is highly tunable with non-magnetic spacers even if the total composition is fixed[40], implying that the ferromagnetism in these two typical MTI materials, $MnBi_2Te_4$-$Bi_2Te_3$ and $(Cr,Bi,Sb)_2Te_3$, are driven by different factors. Combining the observations in Figures 3 and 4 leads to the conclusion that while AFM order is sensitive to the Mn-Mn distance, the FM order is not, but instead, mostly dominated by the overall $Bi_2Te_3$/$MnBi_2Te_4$ ratio. It is worth mentioning that magnetic disorder[46], which can cause a spatial variation in magnetic exchange gap, could become more pronounced with increasing Mn-Mn distance due to reduced interlayer coupling. Accordingly, weakening AFM order with increasing Mn-Mn distance is likely to be a combined result of reduced exchange coupling and enhanced magnetic disorder. On the other hand, considering that magnetic defects tend to promote the FM order as shown in Figure 2c, FM order is less likely to be affected by magnetic disorder in this system. The contrasting response behavior of AFM and FM orders to various effects provide valuable insights into designing novel MTI heterostructure materials.

In summary, we observed coexisting AFM and FM orders in the $MnBi_2Te_4$-$Bi_2Te_3$ heterostructures, which originate from the presence of $Mn_{Bi}$ antisite defects. Through transport measurements, we unveiled that in this system, the increased $Bi_2Te_3$/$MnBi_2Te_4$ ratio suppresses FM order even when the Mn-Mn interlayer distance is kept constant, whereas the AFM order is almost exclusively determined by the Mn-Mn interlayer distance. This observation strongly suggests that in the $MnBi_2Te_4$-$Bi_2Te_3$ system, the AFM order originates from direct interlayer coupling between the neighboring Mn ions whereas the FM order depends only on the overall $Bi_2Te_3$/$MnBi_2Te_4$ ratio regardless of the distance between the $MnBi_2Te_4$ layers. These results reveal the complex interplay between magnetic coupling, Mn-Mn interlayer distance, and $Bi_2Te_3$/$MnBi_2Te_4$ ratio in the $MnBi_2Te_4$-$Bi_2Te_3$ heterostructures. In particular, our work demonstrates that the atomic-layer-engineering can be used as an effective tool to control the magnetic orders in topological $MnBi_2Te_4$-$Bi_2Te_3$ heterostructures, toward novel magneto-topological effects.

**EXPERIMENTAL SECTION**

**Sample Preparation.** We grew all the $MnBi_2Te_4$-$Bi_2Te_3$ heterostructures on $10 \times 10$ mm$^2$ $Al_2O_3$ (0001) substrates using a custom-built MBE system with a base pressure of low $10^{-10}$ Torr. Substrates were treated by methods reported in our previous works[8, 9, 40, 47-49]. The $MnBi_2Te_4$ and $Bi_2Te_3$ layers were deposited at 300 ℃. Between the deposition of each SL $MnBi_2Te_4$ or each QL $Bi_2Te_3$ layer, we annealed the film at 300 ℃ for 1 minute under Te flux. Then Te capping layer was deposited on top after the samples were cooled down to room temperature.

**Transport Measurement.** All measurements were performed using the standard van der Pauw geometry, by manually pressing four indium wires on the corners of each sample. The transport measurements were performed in a Quantum Design Physical Property Measurement System (PPMS; 2 K). Raw data of $R_{xx}$ and $R_{xy}$ were properly symmetrized and anti-symmetrized.

**Computational Methods**. We performed first-principles calculations with the Vienna ab initio simulation package (VASP) based on the density functional theory (DFT)[50-52]. The exchange-correlation functionals are treated by the generalized gradient approximation (GGA) in Perdew-Burke-Ernzerhof (PBE) form[53]. The cutoff energy of plane wave expansion is set to 420 eV. An $18 \times 18 \times 1$ $\Gamma$-center k-point mesh is dense enough for sampling the Brillouin zone. A vacuum space of 15 Å is adopted in periodical direction for avoiding the interactions between adjacent layers. All structures are fully relaxed until Hellmann-Feynman force acting on each atom is less than $10^{-2}$ eV/Å, and the convergence criterion of total energy is set to $10^{-6}$ eV. For describing the strong correlation effects of 3d electrons of Mn, GGA+U method is adopted where the $U_{eff}$ is set to 3 eV. The effect of van der Waals interactions are considered by employing the DFT+D3 method.

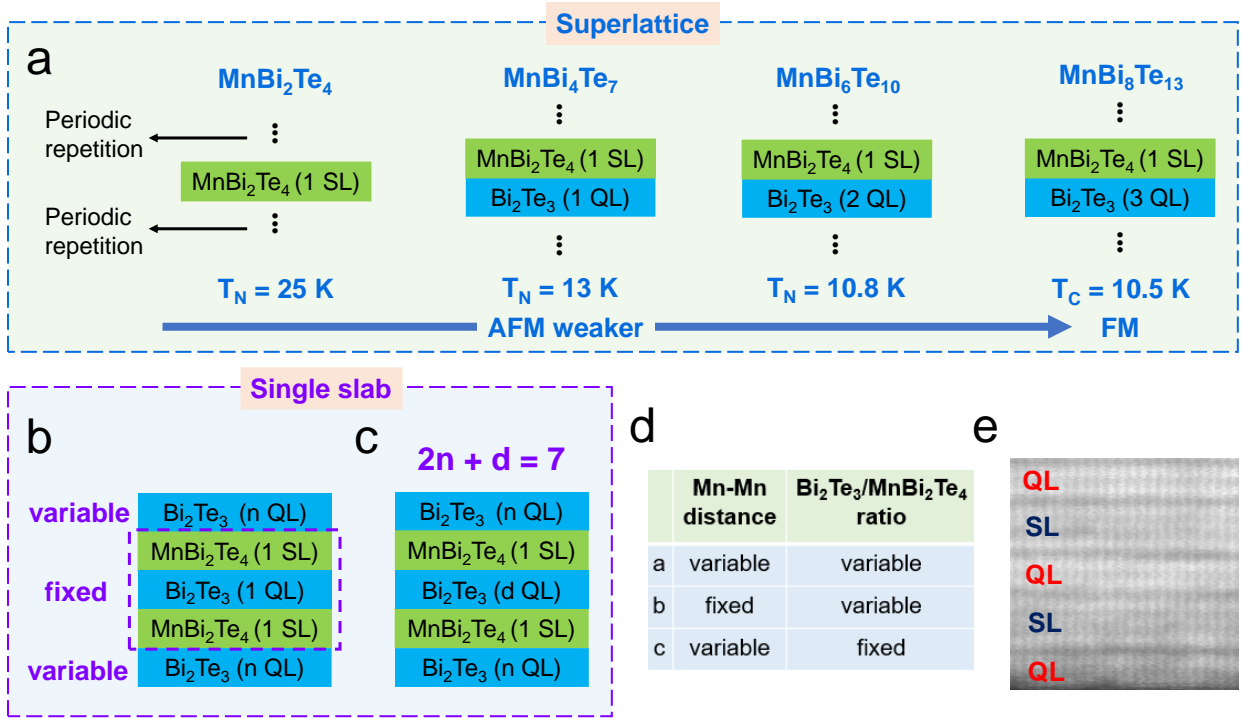

Figure 1. (a) Summary of the studies on $MnBi_2Te_4$-$(Bi_2Te_3)_m$ bulk crystal family, results are extracted from Reference 13-16. (b, c) Illustration of the two sets of $MnBi_2Te_4$-$Bi_2Te_3$ heterostructures used for the current study. In (b) the n QL $Bi_2Te_3$ layers are used as extension layers while keeping the Mn-Mn distance constant in the middle layers, while in (c) the total thickness of $Bi_2Te_3$ is fixed (7 QL) and the spacer $Bi_2Te_3$ layer is varied. (d) Comparison of the three configurations in (a-c). Notably, configuration (a) is superlattice structure while configurations (b) and (c) are single slab heterostructures. (e) High-angle annular dark-field scanning transmission electron microscopy (HAADF-STEM) image of a typical heterostructure sample. The stacking of $MnBi_2Te_4$ and $Bi_2Te_3$ layers are labeled by SL and QL, respectively.

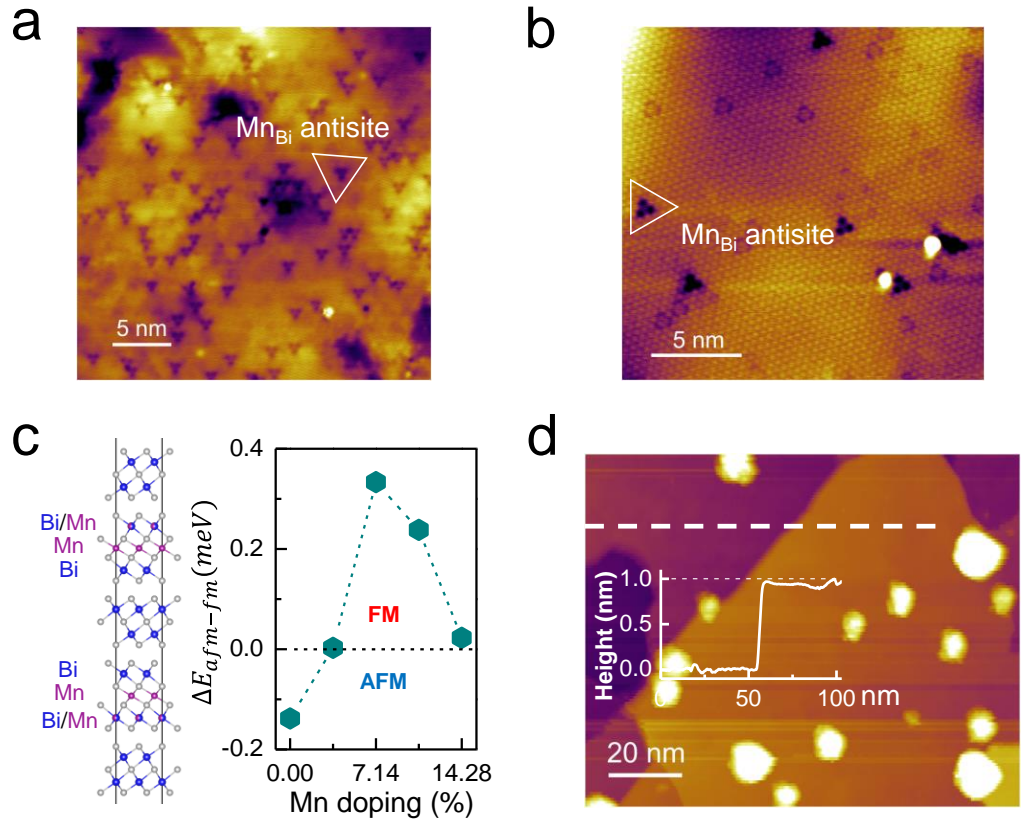

Figure 2. Mn$_{Bi}$ antisite defects toward coexisting FM and AFM orders in the MnBi$_2$Te$_4$-Bi$_2$Te$_3$ heterostructures. (a, b) Scanning tunneling microscopy (STM) images of the Mn$_{Bi}$ antisites collected at the topmost layer of the MnBi$_2$Te$_4$-Bi$_2$Te$_3$ heterostructures (configuration (b)) with (a) n = 0 and (b) n = 4. (c) Left panel is the schematic structure of one possible Mn doping configuration in MnBi$_2$Te$_4$-Bi$_2$Te$_3$ heterostructures. Right panel shows the Mn-doping-level-dependent energy difference between AFM and FM ground states obtained by first principles calculations. (d) The topography image collected on sample n = 4. The inset shows the height profile of terrace with height of around 1 nm, corresponding to the thickness of 1 QL Bi$_2$Te$_3$. The white dots are residual Te from the Te capping layer after decapping process.

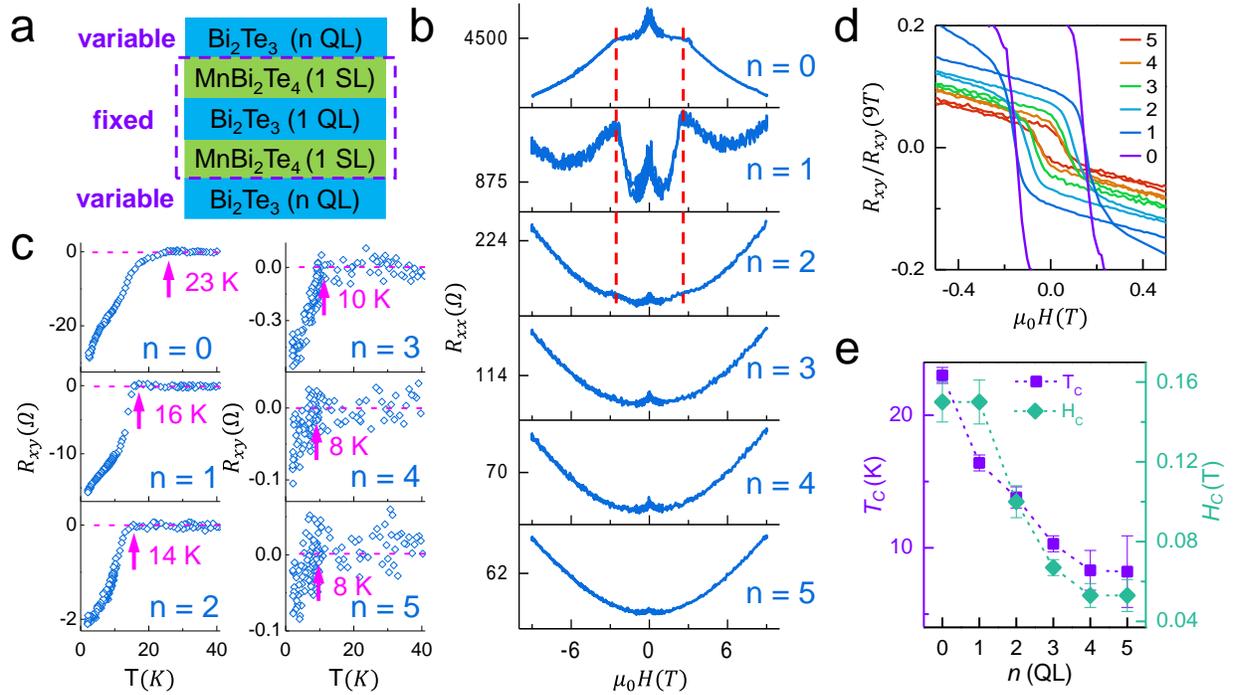

Figure 3. Transport results of the MnBi$_2$Te$_4$-Bi$_2$Te$_3$ heterostructures with configuration (b) (Figure 1b). (a) Schematic illustration of the MnBi$_2$Te$_4$-Bi$_2$Te$_3$ heterostructures with fixed Mn-Mn distance while varying Bi$_2$Te$_3$/MnBi$_2$Te$_4$ ratio. Here "n" refers to the thickness of the outer Bi$_2$Te$_3$ layers. (b) Magnetic field-dependent longitudinal sheet resistance of the MnBi$_2$Te$_4$-Bi$_2$Te$_3$ heterostructures with n values from 0 to 5. (c) Temperature-dependent Hall resistance measured while cooling down under zero magnetic field for all the heterostructure samples. (d) Normalized Hall resistance of the MnBi$_2$Te$_4$-Bi$_2$Te$_3$ heterostructures with n values from 0 to 5 measured at 2 K. (e) Summary of ferromagnetic transition temperatures T$_C$ determined by (c) and coercive fields H$_C$ determined by (d).

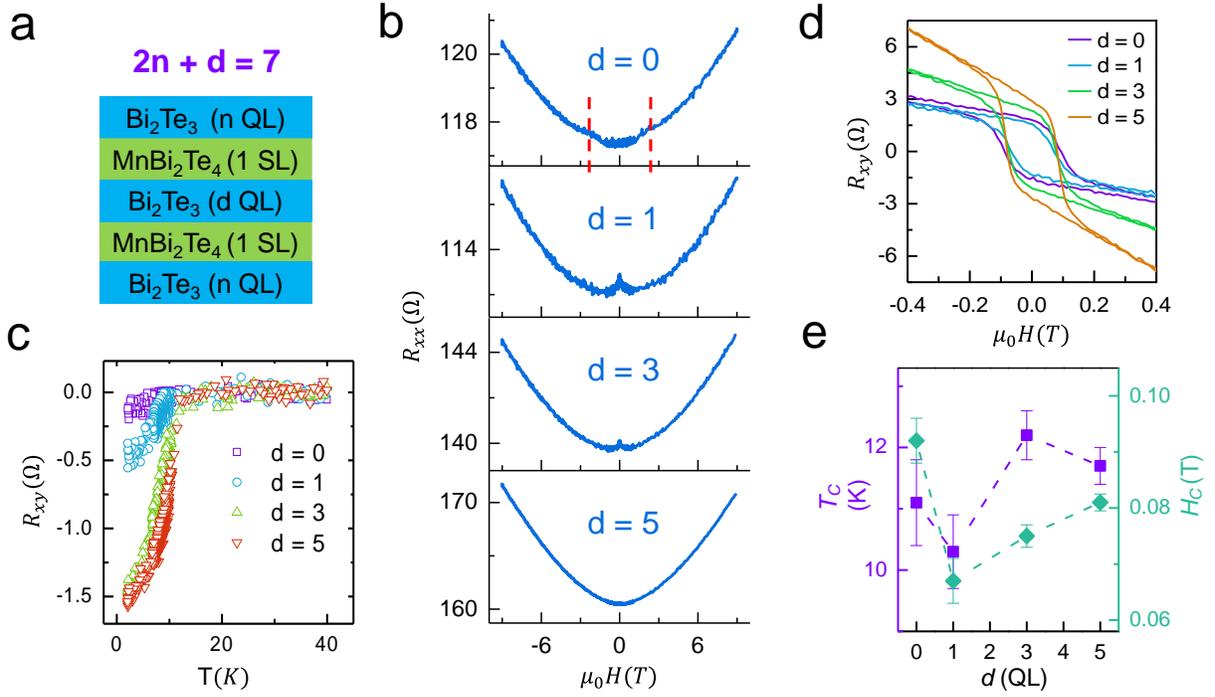

Figure 4. Transport results of the MnBi$_2$Te$_4$-Bi$_2$Te$_3$ heterostructures with configuration (c) (Figure 1c). (a) Schematic illustration of the MnBi$_2$Te$_4$-Bi$_2$Te$_3$ heterostructures with fixed Bi$_2$Te$_3$/MnBi$_2$Te$_4$ ratio while varying Mn-Mn distance. Here "d" refers to the distance between the two MnBi$_2$Te$_4$ layers. (b) Magnetic field-dependent longitudinal sheet resistance of the MnBi$_2$Te$_4$-Bi$_2$Te$_3$ heterostructures with d values from 0 to 5. (c) Hall resistance measured under zero magnetic field for all the heterostructure samples with varying d values. (d) Magnetic-field-dependent Hall resistance of these samples measured at 2 K. (e) Summary of the ferromagnetic transition temperatures and coercive magnetic fields determined from (c) and (d). Note that the shoulder feature at 3 T in the $R_{xx}$ vs $\mu_0 H$ data, characteristic of the AFM order, appears only in the d = 0 sample, confirming that the AFM order strongly depends on the Mn-Mn interlayer distance. In contrast, the strength of the FM order as measured by $H_C$ and $T_C$ in (e) remains little affected by the Mn-Mn distance.

Supporting information:

The RHEED patterns of the $MnBi_2Te_4$-$Bi_2Te_3$ heterostructure, and the longitudinal sheet resistance of the $MnBi_2Te_4$-$Bi_2Te_3$ heterostructures.

AUTHOR INFORMATION

Corresponding Authors

*E-mail: yaoxiong@nimte.ac.cn

*E-mail: hongxin.yang@zju.edu.cn

*E-mail: ohsean@physics.rutgers.edu

Author Contributions

X.Yao and S.O. conceived the experiments; X.Yao, X.Yuan, H.T.Y, D.J., and S.O. grew the thin films; X.Yao performed the transport measurements and analyzed the data with Q.C., H.Y. and S.O.; Q.C. and H.Y. performed the first principles calculations; Z.H. and W.W. performed the STM measurements; X.Yao and S.O. wrote the manuscript with contributions from Q.C. and H.Y.

ACKNOWLEDGMENT

The work at Rutgers is supported by Army Research Office's W911NF2010108, MURI W911NF2020166, and the center for Quantum Materials Synthesis (cQMS), funded by the Gordon and Betty Moore Foundation's EPiQS initiative through grant GBMF10104. The work is also supported by the National Natural Science Foundation of China (Grant No. 12304541) and the Ningbo Science and Technology Bureau (Grant No. 2023J047). The first-principles calculations are supported by the "Pioneer" and "Leading Goose" R&D Program of Zhejiang Province (Grant No. 2022C01053), the National Key Research and Development Program of China (MOST) (Grant No. 2022YFA1405100), the National Natural Science Foundation of China (Grant No. 12174405), Ningbo Key Scientific and Technological Project (Grant No. 2021000215). This research used the Electron Microscopy resources of the Center for Functional Nanomaterials (CFN), which is a U.S. Department of Energy Office of Science User Facility, at Brookhaven National Laboratory under Contract No. DE-SC0012704.

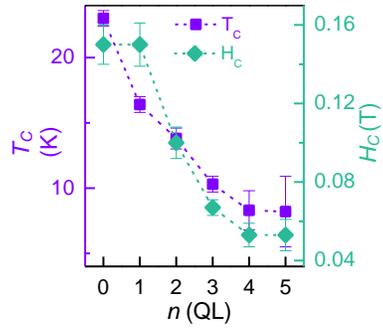

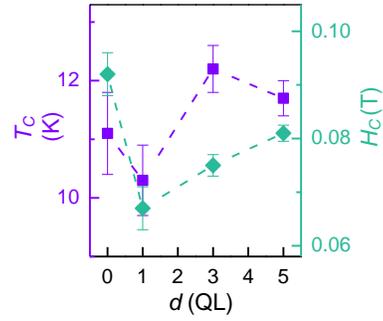

TOC only